\documentstyle[twocolumn,aps,graphicx]{revtex}

\begin{document}
\hyphenation{Ka-pi-tul-nik}

\twocolumn[
\hsize\textwidth\columnwidth\hsize\csname@twocolumnfalse\endcsname

\draft

\title{Superconductor-Insulator Transition in a Capacitively Coupled
Dissipative Environment}

\author{Nadya Mason and Aharon Kapitulnik}
\address{Departments of Applied Physics and of Physics, Stanford
University, Stanford, CA 94305, USA }

\date{\today}

\maketitle

\begin{abstract}
We present results on disordered amorphous films which are
expected to undergo a field-tuned Superconductor-Insulator
Transition. The addition of a parallel ground plane in proximity
to the film changes the character of the transition. Although the
screening effects expected from ``dirty-boson" theories are not
evident, there is evidence that the ground plane couples a certain
type of dissipation into the system, causing a dissipation-induced
phase transition. The dissipation due to the phase transition
couples similarly into quantum phase transition systems such as
superconductor-insulator transitions and Josephson junction
arrays.
\end{abstract}

]

Although quantum phase transitions (QPTs) have attracted much
attention in recent years, there are still discrepancies between
theoretical expectations and experimental results. QPTs occur at
zero temperature, where quantum fluctuations are relevant and the
system's dynamics and thermodynamics are intertwined. A continuous
transition  occurs as one parameter in the Hamiltonian of the
quantum system is   varied; then, quantum critical phenomena are
expected to give rise to interesting, universal physics
\cite{sondhi}. Of  particular interest is the two-dimensional field-tuned
superconductor-insulator transition (SIT), where effects of
Coulomb repulsion, disorder and   dissipation are all expected to
affect the QPT \cite{goldman}. While this transition was  long
thought to be a paradigm of the  ``dirty bosons" theory, recent
experiments demonstrating a  low-temperature metallic state
intervening in the SIT and a coupling   of the system to
dissipation have cast doubt on this model \cite{ephron,mason}.
Similar results on other QPT systems -- especially those expected
to be in the same universality  class, such as the quantum Hall
liquid to insulator transition -- have suggested that the common
simple one-parameter scaling  analysis of QPTs, and concurrent
neglect of dissipation, is no longer sufficient \cite{kapitulnik}.
A better understanding of  the role of dissipation is needed, and
can be achieved if dissipation can be changed in a controlled way.

In this Letter, we present results on the effects of capacitively
coupling a ground plane to an SIT system.  Resistance and
magnetoresistance measurements taken on thin films of
$Mo_{43}Ge_{57}$ insulated by  $\sim 160\dot{A}$ from a thick gold
film manifest two main results: 1)there is no observed change in
physics due to increased screening, and 2) there is evidence of a
dissipation-induced quantum phase transition. The first of these
results contradicts expected theories of the SIT, and is
consistent with what  would be expected in a system where
intrinsic dissipation already exists.  The second result
demonstrates the extent to which dissipation can couple in to the
system. here we demonstrate the fundamental difference between coupling
to dissipation that involves charge exchange and pure capacitive
coupling. Similar results have been found in Josephson junction arrays,
which also undergo dissipative phase transitions
\cite{rimberg}. A strong parallel can thus be made between thin
films and arrays.

The samples used in this study are 30$\AA$ and 40$\AA$ films of
Mo$_{43}$Ge$_{57}$, grown by magnetron sputtering on  SiN
substrates, with a Ge buffer layer and cap. The 30$\dot{A}$ and
40$\dot{A}$ films have sheet resistances at 4.2K of  $R_{N}
\sim$1500 $\Omega / \Box$ and $R_{N} \sim$800 $\Omega / \Box$,
respectively; $T_c$'s of 0.5K and  1K; $H_{C2}$'s  of 1.4T and
1.9T. Previous studies have determined the films to be   amorphous
and homogeneous on  all relevant length scales \cite{graybeal}.
The ``ground plane" samples consisted of MoGe   films patterned
into 4-probe structures, with the  area between voltage taps
covered with an insulating layer (40$\AA$   Ge, 80$\AA$ AlO$_x$,
40$\AA$ Ge), then topped with  a conducting ground plane (20$\AA$
Ti, 400$\AA$ Au). In order to   directly compare samples with and
without a ground  plane, each MoGe film contained three patterns:
one bare sample   ($R_N=1.55k\Omega$), one sample with an
insulating  layer ($R_N=1.5k\Omega$), and one sample with an
insulating layer   covered by a metallic film ($R_N=1.49k\Omega$).
In this way, we were able to determine that the Ge/AlO$_x$
insulating   layer did not affect the behavior of the film, and
that any changes we saw were only due to the gold ground plane.
Also, measurements of the ground plane and bare sample were
normalized to account for the slight differences in $R_N$ at 4K
(due to imprecision in patterning). The films were measured in a
dilution refrigerator using standard low-frequency (${\it
f}_{AC}=27.5Hz$) lock-in techniques with an applied bias of 1nA
(well within the Ohmic regime).

Although, as mentioned previously, the dirty bosons model is not
adequate to fully account for our results, it is still useful to
use this model and its scaling expectations \cite{fisher} as a
guideline for data analysis. This model predicts a field-tuned,
zero-temperature SIT, caused by competition between quantum
fluctuations of the phase of the order  parameter and long-range
Coulomb repulsion. A true superconducting   state is expected to
exist at T=0, when vortices are  localized into a vortex-glass
phase and Cooper pairs are delocalized;   above a critical field,
vortices delocalize while Cooper pairs localize into an insulating
Bose-glass phase. This   SIT is predicted to be continuous, with a
correlation length that diverges as H approaches $H_c$ as
$\xi=\xi_0[(H-H_c)/H_c]^{-\nu}$ with $\xi_0 \sim \xi_{GL}$, the
Ginzburg-Landau correlation length, and $\nu \geq 1$. The
characteristic frequency, $\Omega$, of quantum fluctuations should
vanish near the transition as $\Omega \sim \xi^{-z}$. However,  at
finite temperatures,  thermal energy cuts off this frequency
scale, leading to a temperature-dependent crossover length $L_T
\sim 1/T^{1/z}$. The dynamical exponent, $z$, thus determines the
energy relaxation of the   system, and is expected to have the
value $z=1$  in 2D systems with long-range $1/r$ Coulomb
interactions, and $z=d$ (where $d$ is the dimension) in systems
with  neutral or screened bosons \cite{fisher3,fisher2}. Values of
$\nu$ and $z$ can be extracted from expected scaling forms of the
resistance with temperature, $R \sim
\cal{F}$$_T([H-H_c]/T^{1/z\nu})$, and of the dynamical resistance
with electric field, $\delta V/\delta I \sim
\cal{F}$$_E([H-H_c]/E^{1/(z+1)\nu})$.   Previous measurements of
$\nu$ and $z$ for thin films have   consistently found that  $\nu
\sim 4/3$ and $z \sim 1$ \cite{yazdani,hebard}. The value of
$\nu$=4/3   is consistent with that expected from classical
percolation  and thus adds credence to other indications that the
SIT is a percolative transition \cite{sak}. The value of $z$=1
corresponds to what is expected from theory for a system of
charged bosons. However, recent experiments showing that  the
expected SIT is likely only a crossover to an unusual metallic
state have cast doubt on the dirty-bosons theory,  as well as on
some of its assumptions regarding the critical exponents
\cite{mason,kapitulnik,mason2}. For example, it remained unclear
whether a system already coupled to a bath of fermions could have
Coulomb interactions screened in such a way  that the dynamical
exponent could change from $z$=1 to $z$=2. The relationship
between screening and dissipation was unknown, as were the ways in
which the sample coupled  to dissipation. These questions could be
better answered by studying the effect of a ground plane on SIT
systems.

\begin{figure}
\includegraphics[width=0.95 \columnwidth]{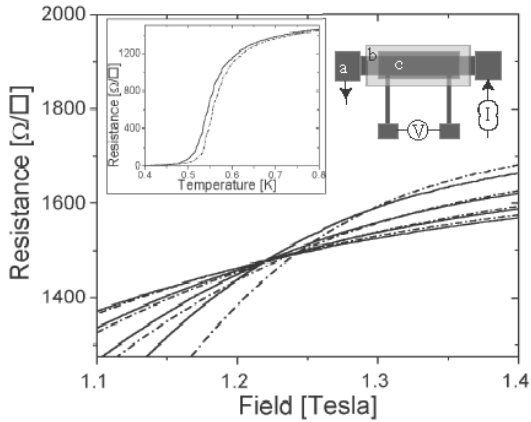}
\caption{Temperature independent ``crossing point" for bare and
ground plane samples, at T=50,100,150, and 200mK. Inset shows H=0
transition for the bare sample (solid line) and the ground plane
sample (dashed line). Top right shows the experimental
configuration: (a) MoGe sample, (b) Insulating-oxide layer,
(c) Gold ground plane.}\label{fig1}
\end{figure}

Fig.\ 1 shows the effect of a ground plane on the  $30\AA$ film
near the critical regions. Although  there is some difference
between the ground plane and bare samples,   measurements of
$T_{c0}$, $R_c$, and $H_c$ show that this  difference is slight.
As can be seen in the inset of Fig.\ 1,   $T_{c0}$ of the ground
plane sample is increased by  $\sim 12$mK, or  $\sim 2.5\%$. $H_c$
and $R_c$, evident in Fig.\ 1 at the temperature   independent
``crossing point" expected from  theory, have similar values with
and without the ground plane (1.24T,   1495$\Omega / \Box$ and
1.22T, 1480$\Omega /  \Box$, respectively). Although the
difference in values near the   critical region is consistent with
increased  screening, it the differences are small. In particular,
neither the universal (e.g. critical exponents) nor the
non-universal (e.g. $R_c$ and $H_c$) properties seem to change
with the proximity of a ground  plane. 

\begin{figure}
\includegraphics[width=0.95 \columnwidth]{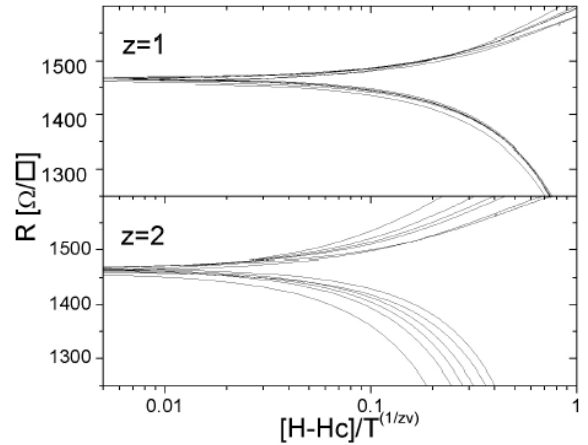}
\caption{Scaling curves for T=30,50,75,100,150,200mK. Upper
graph shows fit for $z\nu$=4/3, lower graph shows fit for
$z\nu$=8/3.} \label{fig2}
\end{figure}

Fig.\ 2 shows the scaling
curves (resistance versus the   scaling parameter,
$[H-H_c]/T^{1/z\nu}$) for the ground  plane sample for values of
$z=1$ and $z=2$ (with $\nu = 4/3$). 
As is evident from the figure,
the  curves scale for $z=1$ much better than they do for $z=2$.
The slight deviations that are evident in the ``z = 1" case appear
at the lowest temperatures, indicating the previously-observed
disruption of scaling at low temperatures \cite{mason}. To confirm
the best value of z, we developed a computer program to
independently fit the scaling curves, using crossing point,
$z\nu$, and data range as independent parameters; from  this, we
determined a best fit for $z\nu$ = 1.25.  This is consistent with
data on dozens of films with different thickness, with or without
ground planes, which show that  $z\nu$ has a value between 1.25
and 1.35. While in plane screening exists at all length scales,
out of plane screening affects the transition only at length
scales below the correlation length; hence, a close ground plane
is expected to change the universality class (i.e. $z=1$ changes
into $z=2$). To observe this effect, the ground plane should be
closer to the SIT system than $\sim \xi_0$. Since $\xi_0 \sim
\xi_{GL} \sim 250 \AA$, our ground plane distance of 160$\AA$
should have been sufficient to see screening in a wide range
around the critical field. It is possible that we do not observe
changes in physics due to screening because there are already
dissipation-causing free fermions in the sample; then, the
addition of a nearby plane of fermions would not cause further
screening.  In this case the value of the critical exponent, $z$,
would be independent of screening effects. It is also possible
that the  mechanism of dissipation (dissipative quantum
tunnelling, for   example) cuts off the correlation length at
short length  scales. In this case, Coulomb interactions could
only be screened at   a very short distance, d, where d$\ll\xi$.

Although the ground plane does not seem to affect the film near
the critical region of the SIT, it does have a dramatic  effect on
the behavior of the system at low temperatures. The main panel of
Fig.\ 3 -- data from a 30$\dot{A}$ film with a ground plane
160$\dot{A}$ -- shows that samples with a ground plane have lower
resistances at low fields and higher resistances at high fields
than samples without a ground plane. The inset is a magnification
of the transition region. 40 $\AA$ thick samples with a ground
plane 250 $\AA$ away from the sample show similar behavior. It was
previously found that bare samples exhibit a metallic phase at low
temperatures, with a levelling of resistance evident as
T$\rightarrow$0 \cite{mason}. The addition of a ground plane seems
to inhibit resistance levelling, or to prevent the system from
entering the metallic phase.

\begin{figure}
\includegraphics[width=0.95 \columnwidth]{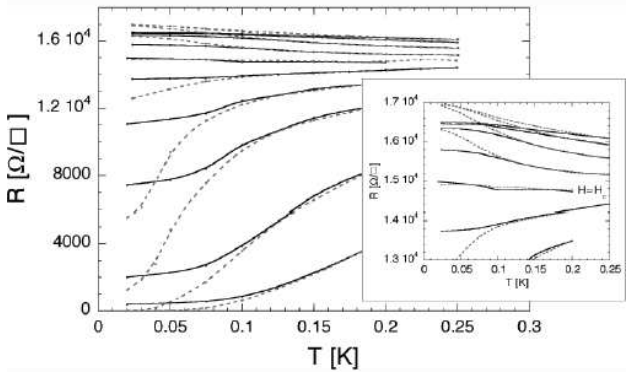}
\caption{Resistance versus temperature curves for bare
(solid line) and ground plane (dashed line) samples. Main figure
shows 30$\dot{A}$ sample at H=0.6,0.8,1.0,1.1,1.18,1.25, and 1.5
Tesla. Inset shows the vicinity of the SIT.} \label{fig3}
\end{figure}

The effect shown in Fig.\ 3 is strongest below $H_c$, where the
proximity of a ground plane clearly promotes superconductivity.
The lower resistance of the  ground plane sample at low
temperatures is too large to be accounted for by the same effects
that caused $T_c$ to change, or by the slight difference in $H_c$
and $R_c$. Rather, the promotion of superconductivity can be
explained by assuming that the capacitive coupling of the ground
plane causes the suppression of the quantum fluctuations that
drive the SIT and destroy superconductivity. A plausible model
involves dissipative coupling \cite{chakravarty}. In this case the
coupling strength, $\alpha$, is assumed to fit a Caldeira-Leggett
model of a dissipative quantum  mechanical environment, where
$\alpha \sim R_q/R$ and $R_q = h/4e^2$ \cite{caldeira}. It has
been proposed that a capacitively coupled ground plane adds Ohmic
dissipation to the system within a fluctuation  frequency range
determined by the capacitance of the ground plane and the
capacitance of the sample \cite{wagenblast}.  Alternatively, the
added capacitance could couple to the inertial term of the phase
fluctuations, adding to its ``mass" \cite{doniach}, which would reduce
the amplitude of the zero-point fluctuations. In
either case, quantum fluctuations are suppressed and the system is
pushed towards the superconducting state, restoring phase order.

A similar promotion of superconductivity due to a coupling to
dissipation was recently observed in arrays of Josephson
junctions. Josephson junctions are expected to undergo an SIT as a
function of the ratio of the Josephson energy to the charging
energy. A dissipation-induced phase transition (DPT) was seen both
when the shunt  resistance of junctions were explicitly changed
\cite{takahide} and when the resistance of a capacitively coupled
plane was varied \cite{rimberg}. The observation of similar DPTs
in both thin films and Josephson junction arrays demonstrates that
dissipation enters into the two systems in a similar  way. The
analogy between the two systems can help identify the origin of
the intervening metallic phase found in SIT systems \cite{mason}.
Because pure capacitive coupling tends to pin a superconducting
phase \cite{schmid,chakravarty}, we conclude that particle
exchange with the heat bath plays a key role in the physics of the
metallic phase. With the suppression of quantum fluctuations, the
motion of the excitations responding to the external field is
diffusive and classical, with characteristics inherited from the
heat bath (which for the thin films is the background residual
fermions, and for arrays the shunts in an RSJ system).  A combined
system of shunt and capacitive coupling therefore demonstrates the
competition between the tendency to pin the superconducting phase
and the tendency to produce a metallic phase at low temperatures.
This conclusion is also backed by a recent theoretical analysis of
a model system of strongly fluctuating superconducting grains
embedded in a metallic background, recently proposed by Feigelman
and Larkin \cite{feigelman} and by Spivak {\it et al.}
\cite{spivak}. In their model, for weak enough inter-grain
coupling,  a low temperature superconducting state is
destroyed by quantum fluctuations, yielding a metallic phase.

Our results on the ``insulating side" of the transition differ
from those on arrays in that our ground
plane also seems to promote the insulating state. As can be seen in
Fig.\ 3,  the ground plane sample has a higher resistance than
the bare sample at low temperatures, for $H > H_c$. The effect is
weaker than in the superconducting state, and more difficult to
analyze because of the weakness of the insulator, and the
proximity of $H_c$ to $H_{c2}$. However, the effect may point to a
restoration of the weakly  localized state in the presence of
strong enough capacitive coupling.

The fact that the ground plane promotes both the superconducting
and insulating states, preventing a levelling of resistance,
suggests that the presence of the ground plane  works {\it
against} the tendency to create a metallic state, strengthening
the influence of the zero-temperature  quantum critical SIT point
to much lower temperatures. In Fig.\ 4 we have marked where the
ground plane and bare sample resistances begin to diverge. Since
this divergence is strongest away from the apparent SIT quantum
critical point (at $H_c$), and non-existent at  $H_c$, this set of
points tracks the region of influence of the quantum critical
point. It is possible that the ground plane manifests a transition
between two distinct  metallic states in the bare sample (one in
the insulating and the other in the superconducting regime). To
verify this observation a stronger insulator will be needed. However, it
is likely that the higher field flattening is a direct consequence of
the sample being only weakly localized, and the true high-field
phase is an insulator.

\begin{figure}
\includegraphics[width=0.95 \columnwidth]{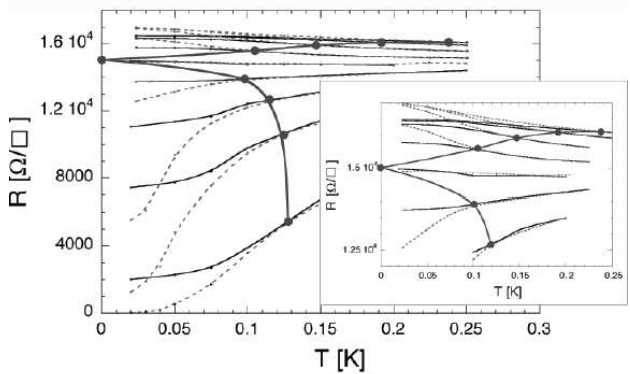}
\caption{Location of the points at which the ground-plane transitions
diverge from the bare-sample transitions for the 30$\dot{A}$ sample. 
Inset shows the vicinity of the SIT.}
\label{fig4}
\end{figure}

In conclusion, we have presented results demonstrating that a
ground plane capacitively coupled to a SIT system which has
inherent dissipation couples to the system in a way that damps
phase fluctuations and promotes a pure SIT. This implies that the
inherent dissipation comes from particle exchange with the heat
bath, which is composed of residual fermions in the film, or of
shunts in an RSJ system. Phase fluctuations can be damped in both
the superconducting and insulating states, leading to a promotion
of either state.\\

We thank David Ephron whose thesis work motivated parts of this study.
We thank Steve Kivelson, Mac Beasley and Seb Doniach for many useful
discussions. Work supported by NSF/DMR. NM thanks Lucent CRFP
Fellowship program for  support.  Samples prepared at Stanford's Center
for Materials Research.

\end{document}